%% file: worm_algorithm_for_lattice_CPN_model.tex
\documentclass{PoS}
\usepackage[utf8]{inputenc}
\usepackage[T1]{fontenc}
\usepackage[pdftex]{graphicx}
\usepackage{tikz}
\usetikzlibrary{arrows,shapes,backgrounds,calc,positioning,patterns,decorations.pathreplacing}
\usepackage{verbatim}
\usetikzlibrary{arrows}
\usepackage{picture}
\usepackage{pslatex}
\usepackage{amsmath}
\usepackage{amsfonts}
\usepackage{amssymb}
\usepackage{cite}
\usepackage{dcolumn}
\usepackage{bm}
\usepackage{bbm}
\usepackage[ngerman,english]{babel}
\usepackage{cancel}
\usepackage{tabularx}
\usepackage{tensor}
\usepackage{mathtools}
\usepackage{float}
\usepackage{fancyhdr}
\usepackage{xfrac}
\usepackage{fontenc}
\usepackage{sidecap}
\usepackage{xcolor}
\usepackage{textpos}
\usepackage{scalerel}
\usepackage{amsmath}

\definecolor{darkblue}{rgb}{0,0,.5}
\definecolor{darkgreen}{rgb}{0,.5,0}
\definecolor{darkred}{rgb}{.5,0,0}
\definecolor{darkyellow}{rgb}{.5,.5,0}
\definecolor{fhl}{rgb}{1.,0,0}
\makeatletter 
\newsavebox\myboxA 
\newsavebox\myboxB 
\newlength\mylenA 
 
\newcommand*\xoverline[2][0.75]{%
    \sbox{\myboxA}{$\m@th#2$}%
    \setbox\myboxB\null
    \ht\myboxB=\ht\myboxA%
    \dp\myboxB=\dp\myboxA%
    \wd\myboxB=#1\wd\myboxA
    \sbox\myboxB{$\m@th\overline{\copy\myboxB}$}
    \setlength\mylenA{\the\wd\myboxA}
    \addtolength\mylenA{-\the\wd\myboxB}%
    \ifdim\wd\myboxB<\wd\myboxA%
       \rlap{\hskip 0.5\mylenA\usebox\myboxB}{\usebox\myboxA}%
    \else 
        \hskip -0.5\mylenA\rlap{\usebox\myboxA}{\hskip 0.5\mylenA\usebox\myboxB}%
    \fi}
\makeatother
\numberwithin{equation}{section}

\let\originalleft\left
\let\originalright\right
\renewcommand{\left}{\mathopen{}\mathclose\bgroup\originalleft}
\renewcommand{\right}{\aftergroup\egroup\originalright}
\newcommand{\e}{\operatorname{e}}
\newcommand{\SU}[1]{\operatorname{SU}\left(#1\right)}
\newcommand{\On}[1]{\operatorname{O}\left(#1\right)}
\newcommand{\Un}[1]{\operatorname{U}\left(#1\right)}
\newcommand{\CP}[1]{\operatorname{CP}\left(#1\right)}

\newcommand{\CPn}[1]{\operatorname{\mathbb{C}P}^{#1}}

\newcommand{\of}[1]{\left(#1\right)}

\newcommand{\roint}[1]{\left[#1\right)}
\newcommand{\bof}[1]{\biggl(\bigg.#1\bigg.\biggr)}
\newcommand{\sof}[1]{\bigl(\big.#1\big.\bigr)}
\newcommand{\ssof}[1]{(#1)}

\newcommand{\cof}[1]{\left\{\right.#1\left.\right\}}
\newcommand{\bcof}[1]{\biggl\{\bigg.#1\bigg.\biggr\}}

\newcommand{\avof}[1]{\left\langle #1\right\rangle}
\newcommand{\savof}[1]{\bigl\langle #1\bigr\rangle}

\newcommand{\ii}{\mathrm{i}}
\newcommand{\idd}[2]{\mathrm{d}^{#2}#1}
\newcommand{\dd}{\mathrm{d}}
\newcommand{\DD}[1]{\mathcal{D}\bigl[#1\bigr]}
\newcommand{\cD}{\mathrm{D}}

\newcommand{\partd}[2]{\frac{\partial #1}{\partial #2}}

\newcommand{\order}[1]{\mathcal{O}\big(#1\big)}

\newcommand{\abs}[1]{\left| #1\right|}
\newcommand{\ssabs}[1]{| #1|}
\newcommand{\sabs}[1]{\big| #1\big|}

\newcommand{\op}[1]{\operatorname{#1}}

\renewcommand*\[{\begin{equation}}
\renewcommand*\]{\end{equation}}
\let\obar\bar
\renewcommand*\bar[1]{\ThisStyle{\xoverline{\SavedStyle #1}}}
\let\ohat\hat
\renewcommand*\hat[1]{\widehat{#1}}
\let\oldstackrel\stackrel
\renewcommand*\stackrel[2]{{\scriptstyle\oldstackrel{#1}{#2}}}

\definecolor{emphcol}{RGB}{0,0,0}
\let\oldemph\emph
\renewcommand*\emph[1]{\oldemph{\textcolor{emphcol}{#1}}}
\DeclareMathSizes{10.95}{9.5}{7}{7} 

\newlength{\hatchspread}
\newlength{\hatchthickness}
\newlength{\hatchshift}
\newcommand{\hatchcolor}{}
\tikzset{hatchspread/.code={\setlength{\hatchspread}{#1}},
         hatchthickness/.code={\setlength{\hatchthickness}{#1}},
         hatchshift/.code={\setlength{\hatchshift}{#1}},
         hatchcolor/.code={\renewcommand{\hatchcolor}{#1}}}
\tikzset{hatchspread=3pt,
         hatchthickness=0.4pt,
         hatchshift=0pt,
         hatchcolor=black}
\pgfdeclarepatternformonly[\hatchspread,\hatchthickness,\hatchshift,\hatchcolor]
   {custom north west lines}
   {\pgfqpoint{\dimexpr-2\hatchthickness}{\dimexpr-2\hatchthickness}}
   {\pgfqpoint{\dimexpr\hatchspread+2\hatchthickness}{\dimexpr\hatchspread+2\hatchthickness}}
   {\pgfqpoint{\dimexpr\hatchspread}{\dimexpr\hatchspread}}
   {
    \pgfsetlinewidth{\hatchthickness}
    \pgfpathmoveto{\pgfqpoint{\dimexpr0pt}{\dimexpr\hatchspread+\hatchshift}}
    \pgfpathlineto{\pgfqpoint{\dimexpr\hatchspread+0.15pt}{\dimexpr\hatchshift-0.15pt}}
    \ifdim \hatchshift > 0pt
      \pgfpathmoveto{\pgfqpoint{\dimexpr-0.15pt}{\dimexpr\hatchshift+0.15pt}}
      \pgfpathlineto{\pgfqpoint{\dimexpr\hatchshift+0.15pt}{\dimexpr-0.15pt}}
    \fi
    \pgfsetstrokecolor{\hatchcolor}
    \pgfusepath{stroke}
   }

\pgfdeclarepatternformonly[\hatchspread,\hatchthickness,\hatchshift,\hatchcolor]
   {custom north east lines}
   {\pgfqpoint{\dimexpr-2\hatchthickness}{\dimexpr-2\hatchthickness}}
   {\pgfqpoint{\dimexpr\hatchspread+2\hatchthickness}{\dimexpr\hatchspread+2\hatchthickness}}
   {\pgfqpoint{\dimexpr\hatchspread}{\dimexpr\hatchspread}}
   {
    \pgfsetlinewidth{\hatchthickness}
    \pgfpathmoveto{\pgfqpoint{0pt}{\dimexpr\hatchshift}}
    \pgfpathlineto{\pgfqpoint{\dimexpr\hatchspread+0.15pt}{\dimexpr\hatchspread+0.15pt+\hatchshift}}
    \ifdim \hatchshift > 0pt
      \pgfpathmoveto{\pgfqpoint{\dimexpr\hatchspread-\hatchshift-0.15pt}{\dimexpr-0.15pt}}
      \pgfpathlineto{\pgfqpoint{\dimexpr\hatchspread+0.15pt}{\dimexpr\hatchshift+0.15pt}}
    \fi
    \pgfsetstrokecolor{\hatchcolor}
    \pgfusepath{stroke}
   }

\pgfdeclarepatternformonly[\hatchspread,\hatchthickness,\hatchshift,\hatchcolor]
   {custom vertical lines}
   {\pgfqpoint{\dimexpr-2\hatchthickness}{\dimexpr-2\hatchthickness}}
   {\pgfqpoint{\dimexpr\hatchspread+2\hatchthickness}{\dimexpr\hatchspread+2\hatchthickness}}
   {\pgfqpoint{\dimexpr\hatchspread}{\dimexpr\hatchspread}}
   {
    \pgfsetlinewidth{\hatchthickness}
    \pgfpathmoveto{\pgfqpoint{\dimexpr\hatchshift}{0pt}}
    \pgfpathlineto{\pgfqpoint{\dimexpr\hatchshift}{\dimexpr\hatchspread+0.15pt}}
    \pgfsetstrokecolor{\hatchcolor}
    \pgfusepath{stroke}
   }

\pgfdeclarepatternformonly[\hatchspread,\hatchthickness,\hatchshift,\hatchcolor]
   {custom horizontal lines}
   {\pgfqpoint{\dimexpr-2\hatchthickness}{\dimexpr-2\hatchthickness}}
   {\pgfqpoint{\dimexpr\hatchspread+2\hatchthickness}{\dimexpr\hatchspread+2\hatchthickness}}
   {\pgfqpoint{\dimexpr\hatchspread}{\dimexpr\hatchspread}}
   {
    \pgfsetlinewidth{\hatchthickness}
    \pgfpathmoveto{\pgfqpoint{0pt}{\dimexpr\hatchshift}}
    \pgfpathlineto{\pgfqpoint{\dimexpr\hatchspread+0.15pt}{\dimexpr\hatchshift}}
    \pgfsetstrokecolor{\hatchcolor}
    \pgfusepath{stroke}
   }

\tikzset{fontscale/.style={font=\relsize{#1}}}
\tikzset{->-/.style={decoration={
  markings,
  mark=at position #1 with {\arrow{>}}},postaction={decorate}}}
\tikzset{-<-/.style={decoration={
  markings,
  mark=at position #1 with {\arrow{<}}},postaction={decorate}}}

\tikzset{cross/.style={cross out,draw,minimum size=2*(#1-\pgflinewidth),inner sep=0pt, outer sep=0pt}}

\title{\begin{textblock*}{100pt}(348pt,-120pt)
\textnormal{\small \texttt{CERN-TH-2017-060}}
\end{textblock*}A Worm Algorithm for the Lattice $\CPn{N-1}$ Model}

\ShortTitle{Worm Algorithm for Lattice $CP(N-1)$ Model}

\author{\speaker{Tobias Rindlisbacher}\\
ETH Z\"urich, Institute for Theoretical Physics,\\ Wolfgang-Pauli-Str. 27, CH - 8093 Z\"urich, Switzerland\footnote{present address:
University of Helsinki, Department of Physics,
P.O. Box 64, FI-00014 University of Helsinki, Finland.}\\
E-mail: \email{tobias.rindlisbacher@helsinki.fi}}

\author{Philippe de Forcrand\\
ETH Z\"urich, Institute for Theoretical Physics,\\ Wolfgang-Pauli-Str. 27, CH - 8093 Z\"urich, Switzerland\\
and\\
CERN, Theory Division, CH-1211 Gen\`eve 23, Switzerland\\
E-mail: \email{forcrand@itp.phys.ethz.ch}}

\abstract{The $\CPn{N-1}$ model in 2D is an interesting toy model for 4D QCD as it possesses confinement, asymptotic freedom and a non-trivial vacuum structure. Due to the lower dimensionality and the absence of fermions, the computational cost for simulating 2D $\CPn{N-1}$ on the lattice is much lower than the one for simulating 4D QCD. However to our knowledge, no efficient algorithm for simulating the lattice $\CPn{N-1}$ model has been tested so far, which also works at finite density. To this end we propose and test a new type of worm algorithm which is appropriate to simulate the lattice $\CPn{N-1}$ model in a dual, flux-variables based representation, in which the introduction of a chemical potential does not give rise to any complications.}

\FullConference{34th annual International Symposium on Lattice Field Theory\\
		24-30 July 2016\\
		University of Southampton, UK}

\begin{document}

\section{Introduction}
The two-dimensional $\CPn{N-1}$ model \cite{Adda,Witten1} has a non-trivial vacuum structure with stable instanton solutions and possesses asymptotic freedom as well as confinement and is therefore an interesting toy-model for four-dimensional $\SU{N}$ Yang-Mills theories.\\
Unfortunately, it turns out that simulations using standard lattice formulation of the model suffer from exponential critical slowing down \cite{Campostrini} of topological modes, which could also not be overcome by the development of cluster \cite{Jansen} and loop algorithms \cite{Wolff}\footnote{In the loop algorithm \cite{Wolff}, the slowing down is caused by a different mechanism (update of long lists) and could possibly be overcome.}.\\
A first worm algorithm, based on a new dual formulation of the $\CPn{N-1}$ model \cite{Chandrasekharan}, has then been proposed and tested in \cite{Vetter}, where it was unfortunately found that the algorithm suffers from an ergodicity problem in relevant parts of parameter space.\\
In this work, we implement ergodic worm algorithms for both, a more recent (and simpler) dual formulation of the $\CPn{N-1}$ partition function \cite{Gattringer1}, as well as for the dual formulation from \cite{Chandrasekharan}, and discuss their performance. This proceeding is based on our full article \cite{RindlisbacherCPN}.

\subsection{The $\CPn{N-1}$ model}\label{ssec:themodel}
The $\CPn{N-1}$ model in $d$ dimensional continuous Euclidean space can be seen as a $\Un{1}$ gauged non-linear $\SU{N}$ sigma model (local $\Un{1}$ and global $\SU{N}$ symmetry)\cite{Cremmer,Witten1,Vecchia2}:
\[
S_{A}\,=\,-\frac{1}{g}\int\idd{x}{d}\,\of{\cD_{\mu}z}^{\dagger}\cdot\of{\cD_{\mu}z}\ ,\label{eq:caction1}
\] 
where $z\in\mathbb{C}^{N}$ is an $N$-component complex scalar field subject to the constraint $z^{\dagger}\,z\,=\,1$, $\cD_{\mu}\,=\,\partial_{\mu}\,+\,\ii\,A_{\mu}$ is the covariant derivative with respect to an auxiliary $\Un{1}$ gauge field $A_{\mu}$ and $g$ is the corresponding coupling strength. By substituting $A_{\mu}\,=\,\frac{\ii}{2}\ssof{z^{\dagger}\cdot\ssof{\partial_{\mu} z}\,-\,\ssof{\partial_{\mu} z^{\dagger}}\cdot z}$, which is the solution to the classical equation of motion for $A_{\mu}$, into \eqref{eq:caction1}, one finds the alternative, \emph{quartic action} \cite{Golo,Witten1,Vecchia2},
\[
S_{Q}\,=\,-\frac{1}{g}\int\idd{x}{d}\,\bof{\ssof{\partial_{\mu}z^{\dagger}}\cdot\ssof{\partial_{\mu}z}\,+\,\frac{1}{4}\sof{z^{\dagger}\cdot\ssof{\partial_{\mu} z}\,-\,\ssof{\partial_{\mu} z^{\dagger}}\cdot{z}}^2}\ ,\label{eq:caction2}
\]
Classically, \eqref{eq:caction1} and \eqref{eq:caction2} are equivalent, but in the quantum theory, quantum fluctuations generate a kinetic term for $A_{\mu}$ in \eqref{eq:caction1} and turn it into a dynamic field \cite{Witten1}.

\subsection{Lattice formulation}\label{ssec:lattform}
Applying the standard discretization strategy to the action $S_{A}$ from \eqref{eq:caction1} one finds
\[
S_{A}\,=\,-\beta\,\sum\limits_{x,\mu}\of{z^{\dagger}\of{x}U_{\mu}\of{x}z\of{x+\hat{\mu}}\,+\,z^{\dagger}\of{x}U^{\dagger}_{\mu}\of{x-\hat{\mu}}z\of{x-\hat{\mu}}\,-\,2}\ ,\label{eq:laction1}
\] 
where $\beta$ is the lattice version of the inverse coupling $\frac{1}{g}$, different lattice sites are labeled by $x$, $\hat{\mu}$ is the vector that points from one lattice site to its nearest neighbor in $\mu$-direction and $U_{\mu}\of{x} \in \Un{1}$ is the parallel transporter with respect to the gauge field $A_{\mu}$ from site $x$ to site $x+\hat{\mu}$ along the corresponding link. The partition function for \eqref{eq:laction1} then reads
\[
Z_{A}\,=\,\int\DD{z^{\dagger},z,U}\,\exp\bof{\beta\,\sum\limits_{x,\mu}\of{z^{\dagger}\of{x}U_{\mu}\of{x}z\of{x+\hat{\mu}}\,+\,z^{\dagger}\of{x}U^{\dagger}_{\mu}\of{x-\hat{\mu}}z\of{x-\hat{\mu}}}}\ ,\label{eq:lpartf1}
\]
where $\DD{z^{\dagger},z,U}=\DD{z^{\dagger},z}\DD{U}$, with $\DD{z^{\dagger},z}=\prod\limits_{x}\delta\sof{\abs{z\of{x}}^2-1}\idd{\bar{z}\of{x}}{N}\,\idd{z\of{x}}{N}$ and $\DD{U}=\prod\limits_{x,\mu}\dd U_{\mu}\of{x}$, where $\dd U_{\mu}\of{x}$ is the $\Un{1}$ Haar measure for the link variable $U_{\mu}\of{x}=\e^{\ii\,\theta_{x,\mu}}$.\\
Discretizing \eqref{eq:caction2}, yields the quartic lattice action and corresponding partition function:\\[-10pt]
\begin{minipage}[t]{0.38\linewidth}
\centering
\[
S_{Q}\,=\,-\beta\,\sum\limits_{x,\mu}\sabs{z^{\dagger}\of{x}\cdot z\of{x+\hat{\mu}}}^2\label{eq:laction2}
\]
\end{minipage}
\hfill%
\begin{minipage}[t]{0.6\linewidth}
\centering
\[
,\quad Z_{Q}\,=\,\int\DD{z^{\dagger},z}\exp\bof{\beta\sum\limits_{x,\mu}\sabs{z^{\dagger}\of{x}\cdot z\of{x+\hat{\mu}}}^2}\ .\label{eq:lpartf2}
\]
\end{minipage}\\[-2pt]

\subsection{Dual formulation}\label{ssec:fluxvar}

To obtain the dual formulation of the partition function \eqref{eq:lpartf2}, we follow the derivations in \cite{Chandrasekharan,Vetter} and start by explicitly writing out all sums in the exponential of \eqref{eq:lpartf2} :
\[
Z_{Q}\,=\,\int\DD{z^{\dagger},z}\,\exp\bof{\beta\,\sum\limits_{x}\sum\limits_{\mu=1}^{d}\sum\limits_{a,b=1}^{N}\sof{\bar{z}_{a}\of{x}z_{b}\of{x}}\sof{\bar{z}_{b}\of{x+\hat{\mu}}z_{a}\of{x+\hat{\mu}}}}\ .\label{eq:partf0}
\]
Now we write the exponential of the summed terms in \eqref{eq:partf0} as the product of exponentials of the individual terms and then use the power series representation for each of these exponentials to find:
\[
Z_{Q}\,=\,\int\DD{z^{\dagger},z}\,\prod\limits_{x}\prod\limits_{\mu=1}^{d}\prod\limits_{a,b=1}^{N}\,\sum\limits_{n_{x,\mu}^{a\,b}=0}^{\infty}\,\bcof{\frac{\beta^{n_{x,\mu}^{a\,b}}}{n_{x,\mu}^{a\,b}!}\of{\sof{\bar{z}_{a}\of{x}z_{b}\of{x}}\sof{\bar{z}_{b}\of{x+\hat{\mu}}z_{a}\of{x+\hat{\mu}}}}^{n_{x,\mu}^{a\,b}}}\ .\label{eq:partf1}
\]
After integrating out the $z_{a}\of{x}$ and defining new integer variables $k_{x,\mu}^{a\,b}=n_{x,\mu}^{a\,b}-n_{x,\mu}^{b\,a}$ and $l_{x,\mu}^{a\,b}=\frac{1}{2}\ssof{n_{x,\mu}^{a\,b}+n_{x,\mu}^{b\,a}-\ssabs{k_{x,\mu}^{a\,b}}}$, so that $k_{x,\mu}^{a\,b}\in\mathbb{Z}$, $l_{x,\mu}^{a\,b}\in\mathbb{N}_{0}$ and $n_{x,\mu}^{a\,b}=\frac{1}{2}\ssof{\ssabs{k_{x,\mu}^{a\,b}}+k_{x,\mu}^{a\,b}}+l_{x,\mu}^{a\,b}$, we find for \eqref{eq:partf1}:
\begin{multline}
Z_{Q}\,=\,\sum\limits_{\cof{k,l}}\prod\limits_{x}\bcof{\bof{\prod\limits_{\mu=1}^{d}\prod\limits_{a,b=1}^{N}\frac{\beta^{\frac{1}{2}\ssof{\ssabs{k_{x,\mu}^{a\,b}}+k_{x,\mu}^{a\,b}}+l_{x,\mu}^{a\,b}}}{\of{\frac{1}{2}\of{\ssabs{k_{x,\mu}^{a\,b}}+k_{x,\mu}^{a\,b}}+l_{x,\mu}^{a\,b}}!}}\\
\frac{\prod\limits_{a}^{N}\bof{\delta\sof{\sum\limits_{\mu=1}^{d}\sum\limits_{b=1}^{N}\sof{k_{x,\mu}^{a\,b}-k_{x-\hat{\mu},\mu}^{a\,b}}}\sof{\sum\limits_{\mu=1}^{d}\sum\limits_{b=1}^{N}\sof{\frac{1}{2}\ssof{\ssabs{k_{x,\mu}^{a\,b}}+\ssabs{k_{x-\hat{\mu},\mu}^{a\,b}}}+l_{x,\mu}^{a\,b}+l_{x-\hat{\mu},\mu}^{a\,b}}}!}}{\sof{N-1+\sum\limits_{\mu=1}^{d}\sum\limits_{c,b=1}^{N}\sof{\frac{1}{2}\ssof{\ssabs{k_{x,\mu}^{c\,b}}+\ssabs{k_{x-\hat{\mu},\mu}^{c\,b}}}+l_{x,\mu}^{c\,b}+l_{x-\hat{\mu},\mu}^{c\,b}}}!}}\ .\label{eq:cpnpartf2}
\end{multline}
In this version of dual $\CPn{N-1}$ partition function, we have in total $N^2$ degrees of freedom per link which are given by the independent components of the $k$- and $l$-variables: the $N\of{N-1}/2$ independent components of the anti-symmetric $k_{x,\mu}^{a\,b}$ are subject to \emph{on-site} constraints, while the $N\of{N+1}/2$ independent components of the symmetric $l_{x,\mu}^{a\,b}$ are unconstrained.\\
The dual version of the auxiliary $\Un{1}$ partition function $Z_{A}$ from \eqref{eq:lpartf1} can be obtained from \eqref{eq:cpnpartf2} by introducing additional link-weight factors (see \cite{RindlisbacherCPN} for more details),
\[
w_{n_{x,\mu}}\of{\beta}=\frac{\beta^{n_{x,\mu}}\e^{-2\,\beta}}{n_{x,\mu}!}\quad,\quad\text{where}\quad n_{x,\mu}=\sum\limits_{a,b=1}^{N}\sof{\tfrac{1}{2}\sabs{k_{x,\mu}^{a\,b}}+l_{x,\mu}^{a\,b}}\ ,\label{eq:addu1weight}
\]
after which one finds
\begin{multline}
Z_{A}\,=\,\sum\limits_{\cof{k,l}}\prod\limits_{x}\bcof{\bof{\prod\limits_{\mu=1}^{d}{\color{red}\frac{\e^{-2\,\beta}}{\sof{\sum\limits_{a,b=1}^{N}\of{\frac{1}{2}\abs{k_{x,\mu}^{a\,b}}+l_{x,\mu}^{a\,b}}}!}}\bof{\prod\limits_{a,b=1}^{N}\frac{\beta^{{\color{red}2}\,\ssof{\frac{1}{2}\ssof{\ssabs{k_{x,\mu}^{a\,b}}+k_{x,\mu}^{a\,b}}+l_{x,\mu}^{a\,b}}}}{\of{\frac{1}{2}\of{\ssabs{k_{x,\mu}^{a\,b}}+k_{x,\mu}^{a\,b}}+l_{x,\mu}^{a\,b}}!}}}\\
\frac{\prod\limits_{a}^{N}\delta\sof{\sum\limits_{\mu=1}^{d}\sum\limits_{b=1}^{N}\sof{k_{x,\mu}^{a\,b}-k_{x-\hat{\mu},\mu}^{a\,b}}}\sof{\sum\limits_{\mu=1}^{d}\sum\limits_{b=1}^{N}\sof{\frac{1}{2}\ssof{\ssabs{k_{x,\mu}^{a\,b}}+\ssabs{k_{x-\hat{\mu},\mu}^{a\,b}}}+l_{x,\mu}^{a\,b}+l_{x-\hat{\mu},\mu}^{a\,b}}}!}{\sof{N-1+\sum\limits_{\mu=1}^{d}\sum\limits_{c,b=1}^{N}\sof{\frac{1}{2}\ssof{\ssabs{k_{x,\mu}^{c\,b}}+\ssabs{k_{x-\hat{\mu},\mu}^{c\,b}}}+l_{x,\mu}^{c\,b}+l_{x-\hat{\mu},\mu}^{c\,b}}}!}}\ ,\label{eq:cpnpartf1}
\end{multline}
where the changes caused by the link weight \eqref{eq:addu1weight} are marked in red and the number of constrained and unconstrained degrees of freedom remains unchanged. We would like to stress that both, \eqref{eq:cpnpartf2} and \eqref{eq:cpnpartf1} can be coupled to $\of{N-1}$ independent chemical potentials without introducing a sign-problem (see \cite{RindlisbacherCPN} for more details).\\
An alternative dual formulation of \eqref{eq:lpartf1}, which relies on only $2\,N$ degrees of freedom per link, and which also remains sign-problem free after the introduction of chemical potentials, has been given in \cite{Gattringer1} and can be written as
\begin{multline}
\tilde{Z}_{A}\,=\,\sum\limits_{\cof{k,\,l}}\bcof{\prod\limits_{x}\bof{\prod\limits_{\nu=1}^{d}\,\e^{-2\,\beta}\delta\sof{\sum\limits_{a=1}^{N} k_{x,\nu}^{a}}\prod\limits_{a=1}^{N}\,\frac{\beta^{\ssabs{k_{x,\nu}^{a}}+2\,l_{x,\nu}^{a}}}{\ssof{\ssabs{k_{x,\nu}^{a}}+l_{x,\nu}^{a}}!\,l_{x,\nu}^{a}!}}\\
\cdot\frac{\prod\limits_{a=1}^{N}\,\delta\sof{\sum\limits_{\nu=1}^{d}\sof{k_{x,\nu}^{a}-k_{x-\hat{\nu},\nu}^{a}}}\sof{\sum\limits_{\nu=1}^{d}\sof{\frac{1}{2}\sof{\ssabs{k_{x,\nu}^{a}}+\ssabs{k_{x-\hat{\nu},\nu}^{a}}}+l_{x,\nu}^{a}+l_{x-\hat{\nu},\nu}^{a}}}!}{\sof{N-1+\sum\limits_{a=1}^{N}\sum\limits_{\nu=1}^{d}\sof{\frac{1}{2}\sof{\ssabs{k_{x,\nu}^{a}}+\ssabs{k_{x-\hat{\nu},\nu}^{a}}}+l_{x,\nu}^{a}+l_{x-\hat{\nu},\nu}^{a}}}!}}\ ,\label{eq:cpnpartf3}
\end{multline}
where we have set the chemical potentials to zero, as we will not discuss finite density here.

\section{The worm algorithm}\label{sec:simtechniques}

\begin{figure}[!h]
\centering
\begin{minipage}[t]{0.49\linewidth}
\centering
\input{tikzimg/constraints.tex}
\caption{Ergodicity problem. For an explanation of the symbols see the legend in Fig.~\ref{fig:worm2}. A change of the $\of{a,b}$-component of the link matrix $k_{x,\nu}$ (change represented by filled and open circles) implies defects (represented by crosses) in the constraints in \eqref{eq:cpnpartf2} for the sites $x$ and $y=x+\ohat{\nu}$ (left-hand grids). By moving the defects from site $y$ around from site to neighboring site, an ordinary worm would update the $\of{a,b}$-components of the $k$ variables that live on the links along which the defects are moved. However, the defects that are caused by the change of the $\of{a,b}$-component of $k_{x,\nu}$ could also be due to a change of an arbitrary sequence $k_{x,\nu}^{a\,c_{1}}\,k_{x,\nu}^{c_{1}\,c_2}\,\cdots\,k_{x,\nu}^{c_n\,b}$ of components of $k_{x,\nu}$, with $n\geq 1$ (example with $n=2$ in right-hand grids), which has to be taken into account.}
\label{fig:worm1}
\end{minipage}
\hfill%
\begin{minipage}[t]{0.49\linewidth}
\centering
\includegraphics[width=\linewidth]{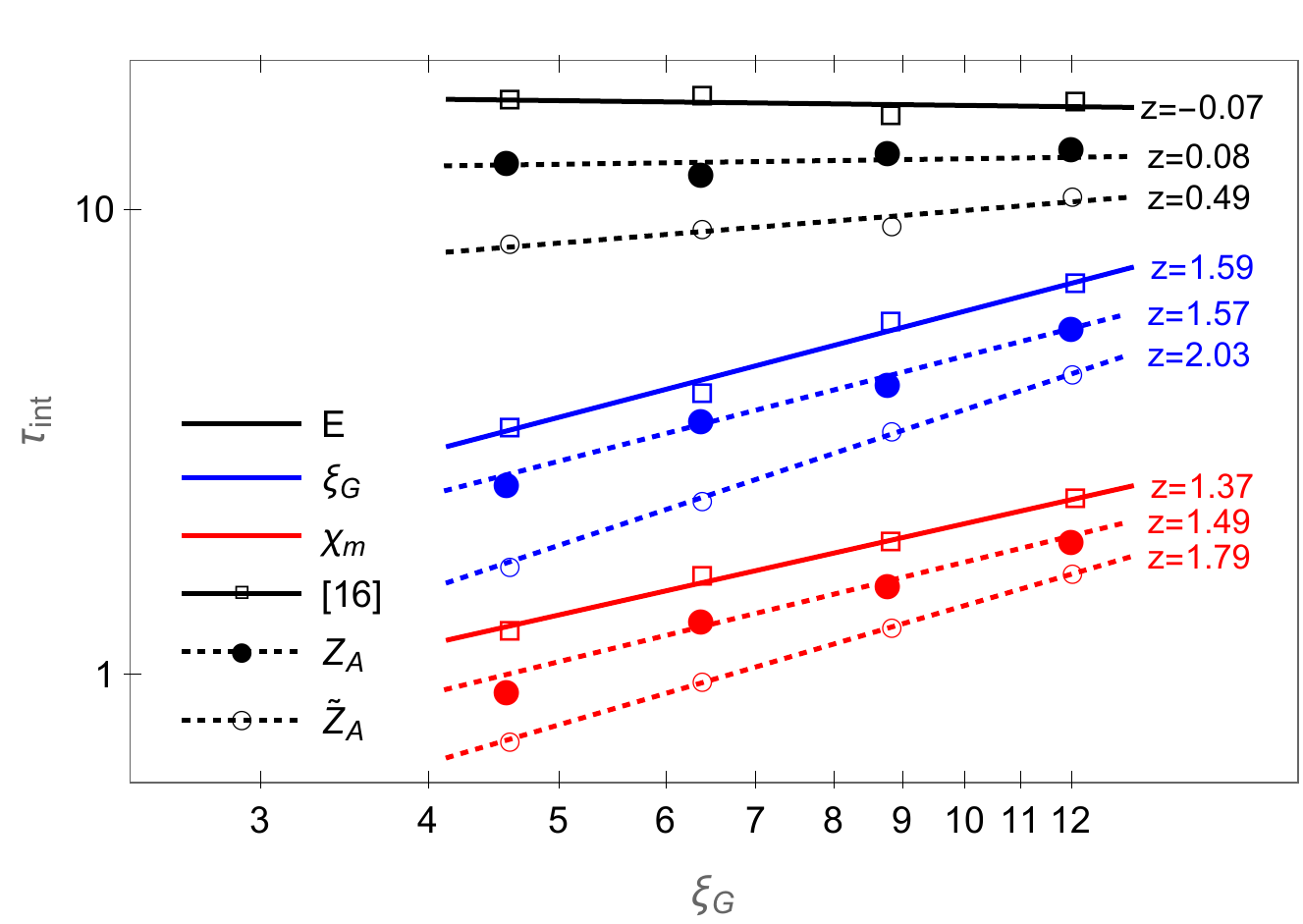}
\caption{\small Log-log plots of $\tau_{int}$ vs. $\xi_{G}$ at fixed $L/\xi_{G}\approx 15$ for the three observables $E$, $\xi_{G}$ and $\chi_{m}$ in the $\CPn{9}$ model, obtained with our two worm algorithms and from data presented in \cite{Flynn}. The $\tau_{int}$ values for the different observables are re-scaled by arbitrary constants, to fit in a common figure. The straight lines correspond to fits of the form $\tau_{int}\propto \xi_{G}^{z}$, where $z$ is the dynamical critical exponent. The filled dots or circles correspond to data for our ISSW algorithm applied to $Z_{A}$ from \eqref{eq:cpnpartf1}, and the open circles to data for an ordinary worm algorithm applied to $\tilde{Z}_{A}$ frim \eqref{eq:cpnpartf3}. The open squares correspond to the data from \cite{Flynn} obtained with an over-heat bath algorithm. All three algorithms show similar behavior.}
  \label{fig:dyncritexp}
\end{minipage}
\end{figure}

The general idea behind a worm algorithm\cite{Prokofev} is to update configuration variables for a partition function $Z$, which are subject to \emph{on-site} constraints as those in \eqref{eq:cpnpartf3} and \eqref{eq:cpnpartf2}, by generating configurations that contribute to some partition functions $Z_{2}\of{x,y}$ instead of $Z$, with $Z_{2}\of{x,y}$ being a \emph{two-point partition function} for the same system that is described by $Z$ itself, but in the presence of an external source at $x$ and an external sink at $y$, which contribute an additional $-1$ or $+1$ to the constraint for the respective site. The constrained configuration variables are then updated one after another by moving the external sink from site to neighboring site and whenever source and sink meet again, a new configuration that contributes to $Z$ can be obtained by removing again the source/sink pair.\\
This strategy works well for the partition function \eqref{eq:cpnpartf3} based on $2\,N$ degrees of freedom per link (see \cite{RindlisbacherCPN}), but for the partition functions \eqref{eq:cpnpartf2} and \eqref{eq:cpnpartf1}, which are based on $N^2$ degrees of freedom per link, it has been found in \cite{Vetter} that an "ordinary" worm algorithm is not sufficiently ergodic to sample the most relevant configurations efficiently. The reason for this ergodicy problem with the ordinary worm is the additional freedom that one has in the $N^2$ d.o.f. per link formulation to compensate for a displacement of a particular external sink (see Fig.~\ref{fig:worm1}). In the remainder of this section we sketch how this additional freedom can be taken into account by generalizing the ordinary worm algorithm to an \emph{internal space sub-worm} algorithm.

\subsection{Internal space sub-worm algorithm}\label{ssec:issworm}
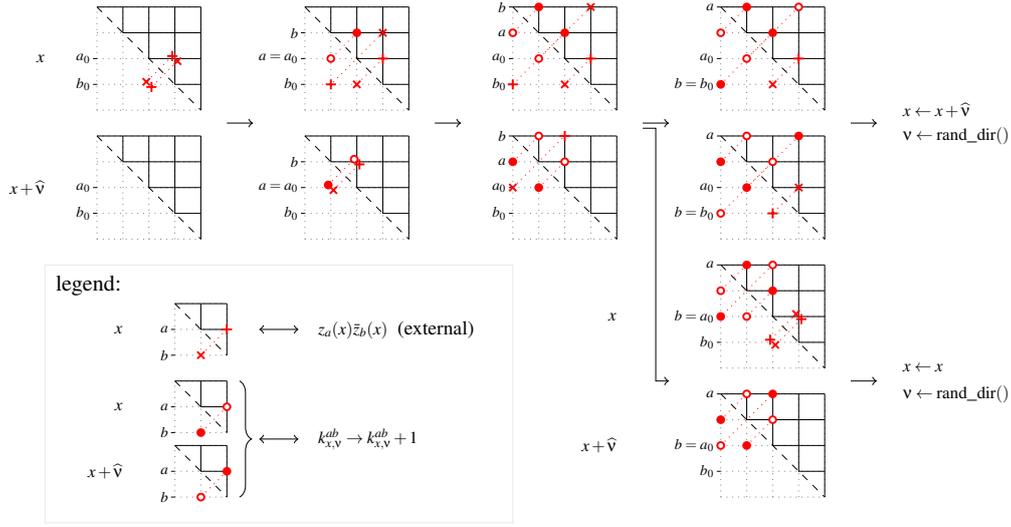
\begin{figure}[!h]
\centering
\begin{minipage}[t]{\linewidth}
\centering
\input{tikzimg/subworm.tex}
\end{minipage}
\caption{\small The figure illustrates (from left to right) how a sub-worm cycle of the ISSW algorithm works, assuming that a positive direction $\nu$ is selected. The small grids represent the sum of all the $k_{x,\mu}$ matrices and sources/sinks that enter the delta-function constraints for $x$ and $x+\ohat{\nu}$ on the second line of \eqref{eq:z2cpnpartf2a}. Note that if a negative direction $\nu$ is chosen at the beginning of the sub-worm cycle, the cycle has to start with $a=b_0$ instead of $a=a_0$ in the second column in the figure, and also for the two possibilities by which the sub-worm cycle can end, as depicted in the last column, the roles of $a_0$ and $b_0$ are interchanged. This is necessary in order to satisfy detailed balance between start and end of the sub-worm cycles (for each possible update, the algorithm must be able to propose the corresponding inverse move as the next update).}
\label{fig:worm2}
\end{figure}

We will discuss the internal space sub-worm (ISSW) algorithm on the example of \eqref{eq:cpnpartf1}, in which case the relevant two-point partition functions, mentioned above, that allow for updates of individual $k$-variables, are given by: 
\begin{multline}
Z_{A,2}^{{\color{red}a_0\,b_0}}{\color{red}\of{x,y}}\,=\,Z_{A}\cdot\avof{\phi^{a_{0}\,b_{0}}\of{x}\phi^{b_{0}\,a_{0}}\of{y}}_{Z_{A}}\,=\,\sum\limits_{\cof{k,l}}\prod\limits_{z}\bcof{\bof{\prod\limits_{\mu=1}^{d}\,\frac{\e^{-2\,\beta}}{\sof{\sum\limits_{a,b=1}^{N}\of{\frac{1}{2}\abs{k_{x,\mu}^{a\,b}}+l_{x,\mu}^{a\,b}}}!}\prod\limits_{a,b=1}^{N}\frac{\beta^{\ssabs{k_{z,\mu}^{a\,b}}+k_{z,\mu}^{a\,b}+2\,l_{z,\mu}^{a\,b}}}{\of{\frac{1}{2}\of{\ssabs{k_{z,\mu}^{a\,b}}+k_{z,\mu}^{a\,b}}+l_{z,\mu}^{a\,b}}!}}\\
\cdot\prod\limits_{a}^{N}\delta\sof{{\color{red}\sof{\delta^{a,b_{0}}-\delta^{a,a_{0}}}\sof{\delta_{x,z}-\delta_{y,z}}}+\sum\limits_{\mu=1}^{d}\sum\limits_{b=1}^{N}\sof{k_{z,\mu}^{a\,b}-k_{z-\hat{\mu},\mu}^{a\,b}}}\\
\cdot\frac{\prod\limits_{a}^{N}\sof{{\color{red}\frac{1}{2}\sof{\delta^{a,a_{0}}+\delta^{a,b_{0}}}\sof{\delta_{x,z}+\delta_{y,z}}}+\sum\limits_{\mu=1}^{d}\sum\limits_{b=1}^{N}\sof{\frac{1}{2}\ssof{\ssabs{k_{z,\mu}^{a\,b}}+\ssabs{k_{z-\hat{\mu},\mu}^{a\,b}}}+l_{z,\mu}^{a\,b}+l_{z-\hat{\mu},\mu}^{a\,b}}}!}{\sof{N-1\,{\color{red}+\,\delta_{x,z}+\delta_{y,z}}+\sum\limits_{\mu=1}^{d}\sum\limits_{c,b=1}^{N}\sof{\frac{1}{2}\ssof{\ssabs{k_{z,\mu}^{c\,b}}+\ssabs{k_{z-\hat{\mu},\mu}^{c\,b}}}+l_{z,\mu}^{c\,b}+l_{z-\hat{\mu},\mu}^{c\,b}}}!}}\ ,\label{eq:z2cpnpartf2a}
\end{multline}
with $\phi^{a b}\of{x}=z_{a}\of{x}\,\bar{z}_{b}\of{x}$. The ISSW algorithm works very similarly to an ordinary worm algorithm, but whenever it is proposed to move the head of the worm from a site $x$ to a neighboring site $x+\hat{\nu}$, a sub-worm cycle as described in Fig.~\ref{fig:worm2} is started instead of simply proposing to update a single $k^{a b}$-variable. For more details see \cite{RindlisbacherCPN}.

\section{Results}\label{sec:results}
We have run extensive tests to verify that our worm and internal space sub-worm algorithms correctly sample configurations for \eqref{eq:cpnpartf3} ($\order{2\,N}$ d.o.f. per link) and \eqref{eq:cpnpartf1} ($\order{N^2}$ d.o.f. per link), respectively, for arbitrary $N$, arbitary system sizes $V$ and in arbitrary dimensions $d$.\\
To test the efficiency of our algorithms, we compare in Fig.~\ref{fig:dyncritexp} for $\CPn{9}$ in $\of{1+1}$ dimensions their dynamical critical exponents $z$ for the integrated auto-correlation times of the average energy and magnetic susceptibility:\\[-15pt]
\begin{minipage}[t]{0.43\linewidth}
\centering
\[
\avof{E}\,=\,-\frac{1}{V}\partd{\log\of{Z}}{\beta}\ \label{eq:avenergy}
\]
\end{minipage}
\hfill%
\begin{minipage}[t]{0.55\linewidth}
\centering
\[
\text{and}\quad \chi_{m}\,=\,\frac{1}{V}\sum\limits_{x,y}\bof{\sum\limits_{a,b}\avof{\phi^{a\,b}\of{x}\phi^{b\,a}\of{y}}-\frac{1}{N}}\ ,\label{eq:magsusc}
\]
\end{minipage}\\[5pt]
and the so-called \emph{second moment correlation length} (see e.g. \cite{Campostrini}),
\[
\xi_{G}\,=\,\frac{1}{2\,\sin\of{\frac{\pi}{L}}}\bof{\tfrac{\sum\limits_{x,y}\sof{\sum\limits_{a,b}\savof{\phi^{a\,b}\of{x}\phi^{b\,a}\of{y}}-\frac{1}{N}}}{\sum\limits_{\bar{x},\bar{y},t_{x},t_{y}}\e^{\frac{2\,\pi\,\ii\,\ssof{t_y-t_x}}{L}}\sof{\sum\limits_{a,b}\savof{\phi^{a\,b}\of{\bar{x},t_{x}}\phi^{b\,a}\of{\bar{y},t_{y}}}-\frac{1}{N}}}-1}^{1/2}\ ,\label{eq:smcorrlen}
\]
with those obtained from data presented in \cite{Flynn} for an over-heat-bath algorithm that directly simulates the (non-dual) system \eqref{eq:lpartf1}. As can be seen, for the given choice of parameters, all three algorithms show very similar dynamical critical exponents.

\section{Conclusion}
We have implemented and tested two worm algorithms which are suitable to simulate the lattice $\CPn{N-1}$ model in two different dual "flux-variable" formulations, which both allow one to couple the model to $\of{N-1}$ independent chemical potentials without causing a sign-problem.\\
For the simulation parameters used in this work ($N=10$, fixed), the worm algorithms do not seem to significantly reduce critical slowing down in comparison to traditional simulation techniques. It is however possible, that the worm algorithms might start to perform better if $N$ is increased, since the slowest dynamical modes are those that change topology, and topological degrees of freedom are integrated out in the dual formulation (see \cite{RindlisbacherCPN}).

\end{document}

%% file: tikzimg/constraints.tex
\scalebox{1.1}{
\begin{tikzpicture}[scale=0.38,nodes={inner sep=0}]
  \pgfpointtransformed{\pgfpointxy{1}{1}};
  \pgfgetlastxy{\vx}{\vy}
  \begin{scope}[node distance=\vx and \vy]
    \def\dxa{9};
    \def\dya{5};
    
    \def\xa{0};
    \def\ya{0};
    \node[left,scale=0.75] at (\xa-0.5,\ya+6-3) {$x$};
    \foreach \i in {1,...,5} {
        \draw [thin,dotted,gray] (\xa+\i,\ya+1) -- (\xa+\i,\ya+5) node[solid,black,above] at (\xa+\i,\ya+5.3) {};
    }
    \foreach \i in {1,...,5} {
        \draw [thin,dotted,gray] (\xa+1,\ya+\i) -- (\xa+5,\ya+\i) node[solid,black,left] at (\xa+0.7,\ya+6-\i) {};
    }
    \foreach \i in {1,...,5} {
      \draw [thin,black!100] (\xa+5,\ya+6-\i) -- (\xa+\i,\ya+6-\i) -- (\xa+\i,\ya+6-1);
    }
    \draw [thin,dashed,black] (\xa+1,\ya+5) -- (\xa+5,\ya+1);

    \def\pxa{4};
    \def\pya{3};
    \node[left,scale=0.65] at (\xa+0.75,\ya+6-\pya) {$a$};
    \draw [thin,black] (\xa+0.85,\ya+6-\pya) -- (\xa+1,\ya+6-\pya);
    \node[left,scale=0.65] at (\xa+0.75,\ya+6-\pxa) {$b$};
    \draw [thin,black] (\xa+0.85,\ya+6-\pxa) -- (\xa+1,\ya+6-\pxa);
    \draw [thin,dotted,red] (\xa+\pxa+0.1,\ya+6-\pya-0.1) -- (\xa+\pya+0.1,\ya+6-\pxa-0.1);
    \draw [thin,dotted,red] (\xa+\pxa-0.1,\ya+6-\pya+0.1) -- (\xa+\pya-0.1,\ya+6-\pxa+0.1);
	\node[draw,thick,circle,inner sep=1,color=red,fill=white] at (\xa+\pxa-0.1,\ya+6-\pya+0.1) {};
	\node[draw,thick,circle,inner sep=1,color=red,fill=red] at (\xa+\pya-0.1,\ya+6-\pxa+0.1) {};
	\node[draw,thick,cross,rotate=45,inner sep=1.5,color=red] at (\xa+\pxa+0.1,\ya+6-\pya-0.1) {};
	\node[draw,thick,cross,inner sep=1.5,color=red] at (\xa+\pya+0.1,\ya+6-\pxa-0.1) {};

    \def\ya{-\dya};
    \node[left,scale=0.75] at (\xa-0.5,\ya+6-3) {$x+\hat{\nu}$};
    \foreach \i in {1,...,5} {
        \draw [thin,dotted,gray] (\xa+\i,\ya+1) -- (\xa+\i,\ya+5) node[solid,black,above] at (\xa+\i,\ya+5.3) {};
    }
    \foreach \i in {1,...,5} {
        \draw [thin,dotted,gray] (\xa+1,\ya+\i) -- (\xa+5,\ya+\i) node[solid,black,left] at (\xa+0.7,\ya+6-\i) {};
    }
    \foreach \i in {1,...,5} {
      \draw [thin,black!100] (\xa+5,\ya+6-\i) -- (\xa+\i,\ya+6-\i) -- (\xa+\i,\ya+6-1);
    }
    \draw [thin,dashed,black] (\xa+1,\ya+5) -- (\xa+5,\ya+1);

    \def\pxa{4};
    \def\pya{3};
    \node[left,scale=0.65] at (\xa+0.75,\ya+6-\pya) {$a$};
    \draw [thin,black] (\xa+0.85,\ya+6-\pya) -- (\xa+1,\ya+6-\pya);
    \node[left,scale=0.65] at (\xa+0.75,\ya+6-\pxa) {$b$};
    \draw [thin,black] (\xa+0.85,\ya+6-\pxa) -- (\xa+1,\ya+6-\pxa);
    \draw [thin,dotted,red] (\xa+\pxa+0.1,\ya+6-\pya-0.1) -- (\xa+\pya+0.1,\ya+6-\pxa-0.1);
    \draw [thin,dotted,red] (\xa+\pxa-0.1,\ya+6-\pya+0.1) -- (\xa+\pya-0.1,\ya+6-\pxa+0.1);
	\node[draw,thick,cross,inner sep=1.5,color=red] at (\xa+\pxa+0.1,\ya+6-\pya-0.1) {};
	\node[draw,thick,cross,rotate=45,inner sep=1.5,color=red] at (\xa+\pya+0.1,\ya+6-\pxa-0.1) {};
	\node[draw,thick,circle,inner sep=1,color=red,fill=red] at (\xa+\pxa-0.1,\ya+6-\pya+0.1) {};
	\node[draw,thick,circle,inner sep=1,color=red,fill=white] at (\xa+\pya-0.1,\ya+6-\pxa+0.1) {};

    \def\xa{\dxa};
    \def\ya{0};
    \node[left,scale=0.75] at (\xa-0.5,\ya+6-3) {$x$};
    \foreach \i in {1,...,5} {
        \draw [thin,dotted,gray] (\xa+\i,\ya+1) -- (\xa+\i,\ya+5) node[solid,black,above] at (\xa+\i,\ya+5.3) {};
    }
    \foreach \i in {1,...,5} {
        \draw [thin,dotted,gray] (\xa+1,\ya+\i) -- (\xa+5,\ya+\i) node[solid,black,left] at (\xa+0.7,\ya+6-\i) {};
    }
    \foreach \i in {1,...,5} {
      \draw [thin,black!100] (\xa+5,\ya+6-\i) -- (\xa+\i,\ya+6-\i) -- (\xa+\i,\ya+6-1);
    }
    \draw [thin,dashed,black] (\xa+1,\ya+5) -- (\xa+5,\ya+1);

    \def\pxa{4};
    \def\pya{3};
    
    \def\pxb{4};
    \def\pyb{1};
    
    \def\pxc{1};
    \def\pyc{3};
    \node[left,scale=0.65] at (\xa+0.75,\ya+6-\pya) {$a$};
    \draw [thin,black] (\xa+0.85,\ya+6-\pya) -- (\xa+1,\ya+6-\pya);
    \node[left,scale=0.65] at (\xa+0.75,\ya+6-\pxa) {$b$};
    \draw [thin,black] (\xa+0.85,\ya+6-\pxa) -- (\xa+1,\ya+6-\pxa);
    \draw [thin,dotted,red] (\xa+\pxa,\ya+6-\pya) -- (\xa+\pya,\ya+6-\pxa);
	\node[draw,thick,cross,rotate=45,inner sep=1.5,color=red] at (\xa+\pxa,\ya+6-\pya) {};
	\node[draw,thick,cross,inner sep=1.5,color=red] at (\xa+\pya,\ya+6-\pxa) {};

    \draw [thin,dotted,red] (\xa+\pxb,\ya+6-\pyb) -- (\xa+\pyb,\ya+6-\pxb);
	\node[draw,thick,circle,inner sep=1,color=red,fill=white] at (\xa+\pxb,\ya+6-\pyb) {};
	\node[draw,thick,circle,inner sep=1,color=red,fill=red] at (\xa+\pyb,\ya+6-\pxb) {};

    \draw [thin,dotted,red] (\xa+\pxc,\ya+6-\pyc) -- (\xa+\pyc,\ya+6-\pxc);
	\node[draw,thick,circle,inner sep=1,color=red,fill=white] at (\xa+\pxc,\ya+6-\pyc) {};
	\node[draw,thick,circle,inner sep=1,color=red,fill=red] at (\xa+\pyc,\ya+6-\pxc) {};

    \def\ya{-\dya};
    \node[left,scale=0.75] at (\xa-0.5,\ya+6-3) {$x+\hat{\nu}$};
    \foreach \i in {1,...,5} {
        \draw [thin,dotted,gray] (\xa+\i,\ya+1) -- (\xa+\i,\ya+5) node[solid,black,above] at (\xa+\i,\ya+5.3) {};
    }
    \foreach \i in {1,...,5} {
        \draw [thin,dotted,gray] (\xa+1,\ya+\i) -- (\xa+5,\ya+\i) node[solid,black,left] at (\xa+0.7,\ya+6-\i) {};
    }
    \foreach \i in {1,...,5} {
      \draw [thin,black!100] (\xa+5,\ya+6-\i) -- (\xa+\i,\ya+6-\i) -- (\xa+\i,\ya+6-1);
    }
    \draw [thin,dashed,black] (\xa+1,\ya+5) -- (\xa+5,\ya+1);

    \def\pxa{3};
    \def\pya{4};
    
    \def\pxb{1};
    \def\pyb{4};
    
    \def\pxc{3};
    \def\pyc{1};
    \node[left,scale=0.65] at (\xa+0.75,\ya+6-\pya) {$a$};
    \draw [thin,black] (\xa+0.85,\ya+6-\pya) -- (\xa+1,\ya+6-\pya);
    \node[left,scale=0.65] at (\xa+0.75,\ya+6-\pxa) {$b$};
    \draw [thin,black] (\xa+0.85,\ya+6-\pxa) -- (\xa+1,\ya+6-\pxa);
    \draw [thin,dotted,red] (\xa+\pxa,\ya+6-\pya) -- (\xa+\pya,\ya+6-\pxa);
	\node[draw,thick,cross,rotate=45,inner sep=1.5,color=red] at (\xa+\pxa,\ya+6-\pya) {};
	\node[draw,thick,cross,inner sep=1.5,color=red] at (\xa+\pya,\ya+6-\pxa) {};

    \draw [thin,dotted,red] (\xa+\pxb,\ya+6-\pyb) -- (\xa+\pyb,\ya+6-\pxb);
	\node[draw,thick,circle,inner sep=1,color=red,fill=white] at (\xa+\pxb,\ya+6-\pyb) {};
	\node[draw,thick,circle,inner sep=1,color=red,fill=red] at (\xa+\pyb,\ya+6-\pxb) {};

    \draw [thin,dotted,red] (\xa+\pxc,\ya+6-\pyc) -- (\xa+\pyc,\ya+6-\pxc);
	\node[draw,thick,circle,inner sep=1,color=red,fill=white] at (\xa+\pxc,\ya+6-\pyc) {};
	\node[draw,thick,circle,inner sep=1,color=red,fill=red] at (\xa+\pyc,\ya+6-\pxc) {};
	
	\node at (0,-6.5) {};

  \end{scope}
\end{tikzpicture}}

%% file: tikzimg/subworm.tex
\scalebox{0.9}{
\begin{tikzpicture}[scale=0.38,nodes={inner sep=0}]
  \pgfpointtransformed{\pgfpointxy{1}{1}};
  \pgfgetlastxy{\vx}{\vy}
  \begin{scope}[node distance=\vx and \vy]
    \def\dxa{8};
    \def\dya{5};
    \def\xa{0};
    \def\ya{\dya};
    \node[left,scale=0.75] at (\xa-1,\ya+6-3) {$x$};
    \foreach \i in {1,...,5} {
        \draw [thin,dotted,gray] (\xa+\i,\ya+1) -- (\xa+\i,\ya+5) node[solid,black,above] at (\xa+\i,\ya+5.3) {};
    }
    \foreach \i in {1,...,5} {
        \draw [thin,dotted,gray] (\xa+1,\ya+\i) -- (\xa+5,\ya+\i) node[solid,black,left] at (\xa+0.7,\ya+6-\i) {};
    }
    \foreach \i in {1,...,5} {
      \draw [thin,black!100] (\xa+5,\ya+6-\i) -- (\xa+\i,\ya+6-\i) -- (\xa+\i,\ya+6-1);
    }
    \draw [thin,dashed,black] (\xa+1,\ya+5) -- (\xa+5,\ya+1);

    \def\pxa{4};
    \def\pya{3};
    \node[left,scale=0.65] at (\xa+0.75,\ya+6-\pya) {$a_{0}$};
    \draw [thin,black] (\xa+0.85,\ya+6-\pya) -- (\xa+1,\ya+6-\pya);
    \node[left,scale=0.65] at (\xa+0.75,\ya+6-\pxa) {$b_{0}$};
    \draw [thin,black] (\xa+0.85,\ya+6-\pxa) -- (\xa+1,\ya+6-\pxa);
    \draw [thin,dotted,red] (\xa+\pxa+0.1,\ya+6-\pya-0.1) -- (\xa+\pya+0.1,\ya+6-\pxa-0.1);
    \draw [thin,dotted,red] (\xa+\pxa-0.1,\ya+6-\pya+0.1) -- (\xa+\pya-0.1,\ya+6-\pxa+0.1);
	\node[draw,thick,cross,inner sep=1.5,color=red] at (\xa+\pxa+0.1,\ya+6-\pya-0.1) {};
	\node[draw,thick,cross,rotate=45,inner sep=1.5,color=red] at (\xa+\pya+0.1,\ya+6-\pxa-0.1) {};
	\node[draw,thick,cross,rotate=45,inner sep=1.5,color=red] at (\xa+\pxa-0.1,\ya+6-\pya+0.1) {};
	\node[draw,thick,cross,inner sep=1.5,color=red] at (\xa+\pya-0.1,\ya+6-\pxa+0.1) {}; 
    
\draw[->,thin,black] (\xa+\dxa-2,\ya+0.5) -- (\xa+\dxa-1,\ya+0.5);

    \def\ya{0};
    \node[left,scale=0.75] at (\xa-1,\ya+6-3) {$x+\hat{\nu}$};
    \foreach \i in {1,...,5} {
        \draw [thin,dotted,gray] (\xa+\i,\ya+1) -- (\xa+\i,\ya+5)  node[solid,black,above] at (\xa+\i,\ya+5.3) {};
    }
    \foreach \i in {1,...,5} {
        \draw [thin,dotted,gray] (\xa+1,\ya+\i) -- (\xa+5,\ya+\i) node[solid,black,left] at (\xa+0.7,\ya+6-\i) {};
    }
    \foreach \i in {1,...,5} {
      \draw [thin,black!100] (\xa+5,\ya+6-\i) -- (\xa+\i,\ya+6-\i) -- (\xa+\i,\ya+6-1);
    }
    \draw [thin,dashed,black] (\xa+1,\ya+5) -- (\xa+5,\ya+1);

    \def\pxa{4};
    \def\pya{3};
    \node[left,scale=0.65] at (\xa+0.75,\ya+6-\pya) {$a_{0}$};
    \draw [thin,black] (\xa+0.85,\ya+6-\pya) -- (\xa+1,\ya+6-\pya);
    \node[left,scale=0.65] at (\xa+0.75,\ya+6-\pxa) {$b_{0}$};
    \draw [thin,black] (\xa+0.85,\ya+6-\pxa) -- (\xa+1,\ya+6-\pxa);      

    \def\xa{\dxa};
    \def\ya{\dya};
    \foreach \i in {1,...,5} {
        \draw [thin,dotted,gray] (\xa+\i,\ya+1) -- (\xa+\i,\ya+5) node[solid,black,above] at (\xa+\i,\ya+5.3) {};
    }
    \foreach \i in {1,...,5} {
        \draw [thin,dotted,gray] (\xa+1,\ya+\i) -- (\xa+5,\ya+\i) node[solid,black,left] at (\xa+0.7,\ya+6-\i) {};
    }
    \foreach \i in {1,...,5} {
      \draw [thin,black!100] (\xa+5,\ya+6-\i) -- (\xa+\i,\ya+6-\i) -- (\xa+\i,\ya+6-1);
    }
    \draw [thin,dashed,black] (\xa+1,\ya+5) -- (\xa+5,\ya+1);

    \def\pxa{4};
    \def\pya{3};
    \def\pxb{4};
    \def\pyb{2};
    \def\pxc{3};
    \def\pyc{2};
    \node[left,scale=0.65] at (\xa+0.75,\ya+6-\pya) {$a=a_{0}$};
    \draw [thin,black] (\xa+0.85,\ya+6-\pya) -- (\xa+1,\ya+6-\pya);
    \node[left,scale=0.65] at (\xa+0.75,\ya+6-\pxa) {$b_{0}$};
    \draw [thin,black] (\xa+0.85,\ya+6-\pxa) -- (\xa+1,\ya+6-\pxa);
    \node[left,scale=0.65] at (\xa+0.75,\ya+6-\pyb) {$b$};
    \draw [thin,black] (\xa+0.85,\ya+6-\pyb) -- (\xa+1,\ya+6-\pyb);
    \draw [thin,dotted,red] (\xa+\pxa,\ya+6-\pya) -- (\xa+\pya,\ya+6-\pxa);
    \draw [thin,dotted,red] (\xa+\pxb,\ya+6-\pyb) -- (\xa+\pyb,\ya+6-\pxb);
    \draw [thin,dotted,red] (\xa+\pxc,\ya+6-\pyc) -- (\xa+\pyc,\ya+6-\pxc);
	\node[draw,thick,cross,rotate=45,inner sep=1.5,color=red] at (\xa+\pxa,\ya+6-\pya) {};
	\node[draw,thick,cross,inner sep=1.5,color=red] at (\xa+\pya,\ya+6-\pxa) {}; 
	\node[draw,thick,cross,inner sep=1.5,color=red] at (\xa+\pxb,\ya+6-\pyb) {};
	\node[draw,thick,cross,rotate=45,inner sep=1.5,color=red] at (\xa+\pyb,\ya+6-\pxb) {};
	\node[draw,thick,circle,inner sep=1.,fill,color=red] at (\xa+\pxc,\ya+6-\pyc) {};
	\node[draw,thick,circle,inner sep=1.,color=red,fill=white] at (\xa+\pyc,\ya+6-\pxc) {};

\draw[->,thin,black] (\xa+\dxa-2,\ya+0.5) -- (\xa+\dxa-1,\ya+0.5);

    \def\ya{0};
    \foreach \i in {1,...,5} {
        \draw [thin,dotted,gray] (\xa+\i,\ya+1) -- (\xa+\i,\ya+5)  node[solid,black,above] at (\xa+\i,\ya+5.3) {};
    }
    \foreach \i in {1,...,5} {
        \draw [thin,dotted,gray] (\xa+1,\ya+\i) -- (\xa+5,\ya+\i) node[solid,black,left] at (\xa+0.7,\ya+6-\i) {};
    }
    \foreach \i in {1,...,5} {
      \draw [thin,black!100] (\xa+5,\ya+6-\i) -- (\xa+\i,\ya+6-\i) -- (\xa+\i,\ya+6-1);
    }
    \draw [thin,dashed,black] (\xa+1,\ya+5) -- (\xa+5,\ya+1);
    
    \def\pxa{4};
    \def\pya{3};
    \def\pxb{2};
    \def\pyb{3};
    \def\pxc{2};
    \def\pyc{3};
    \node[left,scale=0.65] at (\xa+0.75,\ya+6-\pya) {$a=a_{0}$};
    \draw [thin,black] (\xa+0.85,\ya+6-\pya) -- (\xa+1,\ya+6-\pya);
    \node[left,scale=0.65] at (\xa+0.75,\ya+6-\pxa) {$b_{0}$};
    \draw [thin,black] (\xa+0.85,\ya+6-\pxa) -- (\xa+1,\ya+6-\pxa);
    \node[left,scale=0.65] at (\xa+0.75,\ya+6-\pxb) {$b$};
    \draw [thin,black] (\xa+0.85,\ya+6-\pxb) -- (\xa+1,\ya+6-\pxb);
    \draw [thin,dotted,red] (\xa+\pxb+0.1,\ya+6-\pyb-0.1) -- (\xa+\pyb+0.1,\ya+6-\pxb-0.1);
    \draw [thin,dotted,red] (\xa+\pxc-0.1,\ya+6-\pyc+0.1) -- (\xa+\pyc-0.1,\ya+6-\pxc+0.1);
	\node[draw,thick,cross,inner sep=1.5,color=red] at (\xa+\pxb+0.1,\ya+6-\pyb-0.1) {};
	\node[draw,thick,cross,rotate=45,inner sep=1.5,color=red] at (\xa+\pyb+0.1,\ya+6-\pxb-0.1) {};
	\node[draw,thick,circle,inner sep=1.,fill,color=red] at (\xa+\pxc-0.1,\ya+6-\pyc+0.1) {};
	\node[draw,thick,circle,inner sep=1.,color=red,fill=white] at (\xa+\pyc-0.1,\ya+6-\pxc+0.1) {};

    \def\xa{2*\dxa};
    \def\ya{\dya};
    \foreach \i in {1,...,5} {
        \draw [thin,dotted,gray] (\xa+\i,\ya+1) -- (\xa+\i,\ya+5) node[solid,black,above] at (\xa+\i,\ya+5.3) {};
    }
    \foreach \i in {1,...,5} {
        \draw [thin,dotted,gray] (\xa+1,\ya+\i) -- (\xa+5,\ya+\i) node[solid,black,left] at (\xa+0.7,\ya+6-\i) {};
    }
    \foreach \i in {1,...,5} {
      \draw [thin,black!100] (\xa+5,\ya+6-\i) -- (\xa+\i,\ya+6-\i) -- (\xa+\i,\ya+6-1);
    }
    \draw [thin,dashed,black] (\xa+1,\ya+5) -- (\xa+5,\ya+1);
    \def\pxa{4};
    \def\pya{3};
    \def\pxb{4};
    \def\pyb{1};
    \def\pxc{3};
    \def\pyc{2};
    \def\pxd{2};
    \def\pyd{1};
    \node[left,scale=0.65] at (\xa+0.75,\ya+6-\pya) {$a_{0}$};
    \draw [thin,black] (\xa+0.85,\ya+6-\pya) -- (\xa+1,\ya+6-\pya);
    \node[left,scale=0.65] at (\xa+0.75,\ya+6-\pxa) {$b_{0}$};
    \draw [thin,black] (\xa+0.85,\ya+6-\pxa) -- (\xa+1,\ya+6-\pxa);
    \node[left,scale=0.65] at (\xa+0.75,\ya+6-\pyc) {$a$};
    \draw [thin,black] (\xa+0.85,\ya+6-\pyc) -- (\xa+1,\ya+6-\pyc);
    \node[left,scale=0.65] at (\xa+0.75,\ya+6-\pyb) {$b$};
    \draw [thin,black] (\xa+0.85,\ya+6-\pyb) -- (\xa+1,\ya+6-\pyb);
    \draw [thin,dotted,red] (\xa+\pxa,\ya+6-\pya) -- (\xa+\pya,\ya+6-\pxa);
    \draw [thin,dotted,red] (\xa+\pxb,\ya+6-\pyb) -- (\xa+\pyb,\ya+6-\pxb);
    \draw [thin,dotted,red] (\xa+\pxc,\ya+6-\pyc) -- (\xa+\pyc,\ya+6-\pxc);
    \draw [thin,dotted,red] (\xa+\pxd,\ya+6-\pyd) -- (\xa+\pyd,\ya+6-\pxd);
	\node[draw,thick,cross,rotate=45,inner sep=1.5,color=red] at (\xa+\pxa,\ya+6-\pya) {};
	\node[draw,thick,cross,inner sep=1.5,color=red] at (\xa+\pya,\ya+6-\pxa) {}; 
	\node[draw,thick,cross,inner sep=1.5,color=red] at (\xa+\pxb,\ya+6-\pyb) {};
	\node[draw,thick,cross,rotate=45,inner sep=1.5,color=red] at (\xa+\pyb,\ya+6-\pxb) {};
	\node[draw,thick,circle,inner sep=1.,fill,color=red] at (\xa+\pxc,\ya+6-\pyc) {};
	\node[draw,thick,circle,inner sep=1.,color=red,fill=white] at (\xa+\pyc,\ya+6-\pxc) {};
	\node[draw,thick,circle,inner sep=1.,fill,color=red] at (\xa+\pxd,\ya+6-\pyd) {};
	\node[draw,thick,circle,inner sep=1.,color=red,fill=white] at (\xa+\pyd,\ya+6-\pxd) {};

\draw[->,thin,black] (\xa+\dxa-2,\ya+0.5) -- (\xa+\dxa-1,\ya+0.5);
\draw[->,thin,black] (\xa+\dxa-2,\ya+0.5-0.2) -- (\xa+\dxa-2+0.5,\ya+0.5-0.2) --(\xa+\dxa-2+0.5,\ya+0.5-2*\dya) -- (\xa+\dxa-1,\ya+0.5-2*\dya);

    \def\ya{0};
    \foreach \i in {1,...,5} {
        \draw [thin,dotted,gray] (\xa+\i,\ya+1) -- (\xa+\i,\ya+5)  node[solid,black,above] at (\xa+\i,\ya+5.3) {};
    }
    \foreach \i in {1,...,5} {
        \draw [thin,dotted,gray] (\xa+1,\ya+\i) -- (\xa+5,\ya+\i) node[solid,black,left] at (\xa+0.7,\ya+6-\i) {};
    }
    \foreach \i in {1,...,5} {
      \draw [thin,black!100] (\xa+5,\ya+6-\i) -- (\xa+\i,\ya+6-\i) -- (\xa+\i,\ya+6-1);
    }
    \draw [thin,dashed,black] (\xa+1,\ya+5) -- (\xa+5,\ya+1);
    \def\pxa{4};
    \def\pya{3};
    \def\pxb{1};
    \def\pyb{3};
    \def\pxc{2};
    \def\pyc{3};
    \def\pxd{1};
    \def\pyd{2};
    \node[left,scale=0.65] at (\xa+0.75,\ya+6-\pya) {$a_{0}$};
    \draw [thin,black] (\xa+0.85,\ya+6-\pya) -- (\xa+1,\ya+6-\pya);
    \node[left,scale=0.65] at (\xa+0.75,\ya+6-\pxa) {$b_{0}$};
    \draw [thin,black] (\xa+0.85,\ya+6-\pxa) -- (\xa+1,\ya+6-\pxa);
    \node[left,scale=0.65] at (\xa+0.75,\ya+6-\pxc) {$a$};
    \draw [thin,black] (\xa+0.85,\ya+6-\pxc) -- (\xa+1,\ya+6-\pxc);
    \node[left,scale=0.65] at (\xa+0.75,\ya+6-\pxb) {$b$};
    \draw [thin,black] (\xa+0.85,\ya+6-\pxb) -- (\xa+1,\ya+6-\pxb);
    \draw [thin,dotted,red] (\xa+\pxb,\ya+6-\pyb) -- (\xa+\pyb,\ya+6-\pxb);
    \draw [thin,dotted,red] (\xa+\pxc,\ya+6-\pyc) -- (\xa+\pyc,\ya+6-\pxc);
    \draw [thin,dotted,red] (\xa+\pxd,\ya+6-\pyd) -- (\xa+\pyd,\ya+6-\pxd);
	\node[draw,thick,cross,inner sep=1.5,color=red] at (\xa+\pxb,\ya+6-\pyb) {};
	\node[draw,thick,cross,rotate=45,inner sep=1.5,color=red] at (\xa+\pyb,\ya+6-\pxb) {};
	\node[draw,thick,circle,inner sep=1.,fill,color=red] at (\xa+\pxc,\ya+6-\pyc) {};
	\node[draw,thick,circle,inner sep=1.,color=red,fill=white] at (\xa+\pyc,\ya+6-\pxc) {};
	\node[draw,thick,circle,inner sep=1.,fill,color=red] at (\xa+\pxd,\ya+6-\pyd) {};
	\node[draw,thick,circle,inner sep=1.,color=red,fill=white] at (\xa+\pyd,\ya+6-\pxd) {};

    \def\xa{3*\dxa};
    \def\ya{\dya};
    \foreach \i in {1,...,5} {
        \draw [thin,dotted,gray] (\xa+\i,\ya+1) -- (\xa+\i,\ya+5) node[solid,black,above] at (\xa+\i,\ya+5.3) {};
    }
    \foreach \i in {1,...,5} {
        \draw [thin,dotted,gray] (\xa+1,\ya+\i) -- (\xa+5,\ya+\i) node[solid,black,left] at (\xa+0.7,\ya+6-\i) {};
    }
    \foreach \i in {1,...,5} {
      \draw [thin,black!100] (\xa+5,\ya+6-\i) -- (\xa+\i,\ya+6-\i) -- (\xa+\i,\ya+6-1);
    }
    \draw [thin,dashed,black] (\xa+1,\ya+5) -- (\xa+5,\ya+1);
    \def\pxa{4};
    \def\pya{3};
    \def\pxb{4};
    \def\pyb{1};
    \def\pxc{3};
    \def\pyc{2};
    \def\pxd{2};
    \def\pyd{1};
    \node[left,scale=0.65] at (\xa+0.75,\ya+6-\pya) {$a_{0}$};
    \draw [thin,black] (\xa+0.85,\ya+6-\pya) -- (\xa+1,\ya+6-\pya);
    \node[left,scale=0.65] at (\xa+0.75,\ya+6-\pxa) {$b=b_{0}$};
    \draw [thin,black] (\xa+0.85,\ya+6-\pxa) -- (\xa+1,\ya+6-\pxa);
    \node[left,scale=0.65] at (\xa+0.75,\ya+6-\pyb) {$a$};
    \draw [thin,black] (\xa+0.85,\ya+6-\pyb) -- (\xa+1,\ya+6-\pyb);
    \draw [thin,dotted,red] (\xa+\pxa,\ya+6-\pya) -- (\xa+\pya,\ya+6-\pxa);
    \draw [thin,dotted,red] (\xa+\pxb,\ya+6-\pyb) -- (\xa+\pyb,\ya+6-\pxb);
    \draw [thin,dotted,red] (\xa+\pxc,\ya+6-\pyc) -- (\xa+\pyc,\ya+6-\pxc);
    \draw [thin,dotted,red] (\xa+\pxd,\ya+6-\pyd) -- (\xa+\pyd,\ya+6-\pxd);
	\node[draw,thick,cross,rotate=45,inner sep=1.5,color=red] at (\xa+\pxa,\ya+6-\pya) {};
	\node[draw,thick,cross,inner sep=1.5,color=red] at (\xa+\pya,\ya+6-\pxa) {}; 
	\node[draw,thick,circle,inner sep=1.,color=red,fill=white] at (\xa+\pxb,\ya+6-\pyb) {};
	\node[draw,thick,circle,inner sep=1.,fill,color=red] at (\xa+\pyb,\ya+6-\pxb) {};
	\node[draw,thick,circle,inner sep=1.,fill,color=red] at (\xa+\pxc,\ya+6-\pyc) {};
	\node[draw,thick,circle,inner sep=1.,color=red,fill=white] at (\xa+\pyc,\ya+6-\pxc) {};
	\node[draw,thick,circle,inner sep=1.,fill,color=red] at (\xa+\pxd,\ya+6-\pyd) {};
	\node[draw,thick,circle,inner sep=1.,color=red,fill=white] at (\xa+\pyd,\ya+6-\pxd) {};
	
	\draw[->,thin,black] (\xa+\dxa-2,\ya+0.5) -- (\xa+\dxa-1,\ya+0.5);
	\node[black,right,scale=0.75] at (\xa+\dxa,\ya+1) {$x\leftarrow x+\hat{\nu}$};
	\node[black,right,scale=0.75] at (\xa+\dxa,\ya) {$\nu\leftarrow \op{rand\_dir}\of{}$};
	
    \def\ya{0};
    \foreach \i in {1,...,5} {
        \draw [thin,dotted,gray] (\xa+\i,\ya+1) -- (\xa+\i,\ya+5)  node[solid,black,above] at (\xa+\i,\ya+5.3) {};
    }
    \foreach \i in {1,...,5} {
        \draw [thin,dotted,gray] (\xa+1,\ya+\i) -- (\xa+5,\ya+\i) node[solid,black,left] at (\xa+0.7,\ya+6-\i) {};
    }
    \foreach \i in {1,...,5} {
      \draw [thin,black!100] (\xa+5,\ya+6-\i) -- (\xa+\i,\ya+6-\i) -- (\xa+\i,\ya+6-1);
    }
    \draw [thin,dashed,black] (\xa+1,\ya+5) -- (\xa+5,\ya+1);
    \def\pxa{4};
    \def\pya{3};
    \def\pxb{1};
    \def\pyb{4};
    \def\pxc{2};
    \def\pyc{3};
    \def\pxd{1};
    \def\pyd{2};
    \node[left,scale=0.65] at (\xa+0.75,\ya+6-\pya) {$a_{0}$};
    \draw [thin,black] (\xa+0.85,\ya+6-\pya) -- (\xa+1,\ya+6-\pya);
    \node[left,scale=0.65] at (\xa+0.75,\ya+6-\pxa) {$b=b_{0}$};
    \draw [thin,black] (\xa+0.85,\ya+6-\pxa) -- (\xa+1,\ya+6-\pxa);
    \node[left,scale=0.65] at (\xa+0.75,\ya+6-\pxb) {$a$};
    \draw [thin,black] (\xa+0.85,\ya+6-\pxb) -- (\xa+1,\ya+6-\pxb);
    \draw [thin,dotted,red] (\xa+\pxa,\ya+6-\pya) -- (\xa+\pya,\ya+6-\pxa);
    \draw [thin,dotted,red] (\xa+\pxb,\ya+6-\pyb) -- (\xa+\pyb,\ya+6-\pxb);
    \draw [thin,dotted,red] (\xa+\pxc,\ya+6-\pyc) -- (\xa+\pyc,\ya+6-\pxc);
    \draw [thin,dotted,red] (\xa+\pxd,\ya+6-\pyd) -- (\xa+\pyd,\ya+6-\pxd);
	\node[draw,thick,cross,inner sep=1.5,color=red] at (\xa+\pxa,\ya+6-\pya) {};
	\node[draw,thick,cross,rotate=45,inner sep=1.5,color=red] at (\xa+\pya,\ya+6-\pxa) {}; 
	\node[draw,thick,circle,inner sep=1.,color=red,fill=white] at (\xa+\pxb,\ya+6-\pyb) {};
	\node[draw,thick,circle,inner sep=1.,fill,color=red] at (\xa+\pyb,\ya+6-\pxb) {};
	\node[draw,thick,circle,inner sep=1.,fill,color=red] at (\xa+\pxc,\ya+6-\pyc) {};
	\node[draw,thick,circle,inner sep=1.,color=red,fill=white] at (\xa+\pyc,\ya+6-\pxc) {};
	\node[draw,thick,circle,inner sep=1.,fill,color=red] at (\xa+\pxd,\ya+6-\pyd) {};
	\node[draw,thick,circle,inner sep=1.,color=red,fill=white] at (\xa+\pyd,\ya+6-\pxd) {};

    \def\xa{3*\dxa};
    \def\ya{-\dya};

    \node[left,scale=0.75] at (\xa-3,\ya+6-3) {$x$};    
    
    \foreach \i in {1,...,5} {
        \draw [thin,dotted,gray] (\xa+\i,\ya+1) -- (\xa+\i,\ya+5) node[solid,black,above] at (\xa+\i,\ya+5.3) {};
    }
    \foreach \i in {1,...,5} {
        \draw [thin,dotted,gray] (\xa+1,\ya+\i) -- (\xa+5,\ya+\i) node[solid,black,left] at (\xa+0.7,\ya+6-\i) {};
    }
    \foreach \i in {1,...,5} {
      \draw [thin,black!100] (\xa+5,\ya+6-\i) -- (\xa+\i,\ya+6-\i) -- (\xa+\i,\ya+6-1);
    }
    \draw [thin,dashed,black] (\xa+1,\ya+5) -- (\xa+5,\ya+1);
    \def\pxa{4};
    \def\pya{3};
    \def\pxb{3};
    \def\pyb{1};
    \def\pxc{3};
    \def\pyc{2};
    \def\pxd{2};
    \def\pyd{1};
    
    \node[left,scale=0.65] at (\xa+0.75,\ya+6-\pya) {$b=a_{0}$};
    \draw [thin,black] (\xa+0.85,\ya+6-\pya) -- (\xa+1,\ya+6-\pya);
    \node[left,scale=0.65] at (\xa+0.75,\ya+6-\pxa) {$b_{0}$};
    \draw [thin,black] (\xa+0.85,\ya+6-\pxa) -- (\xa+1,\ya+6-\pxa);
    \node[left,scale=0.65] at (\xa+0.75,\ya+6-\pyb) {$a$};
    \draw [thin,black] (\xa+0.85,\ya+6-\pyb) -- (\xa+1,\ya+6-\pyb);
    \draw [thin,dotted,red] (\xa+\pxa+0.1,\ya+6-\pya-0.1) -- (\xa+\pya+0.1,\ya+6-\pxa-0.1);
    \draw [thin,dotted,red] (\xa+\pxa-0.1,\ya+6-\pya+0.1) -- (\xa+\pya-0.1,\ya+6-\pxa+0.1);    
    
    \draw [thin,dotted,red] (\xa+\pxb,\ya+6-\pyb) -- (\xa+\pyb,\ya+6-\pxb);
    \draw [thin,dotted,red] (\xa+\pxc,\ya+6-\pyc) -- (\xa+\pyc,\ya+6-\pxc);
    \draw [thin,dotted,red] (\xa+\pxd,\ya+6-\pyd) -- (\xa+\pyd,\ya+6-\pxd);
	\node[draw,thick,cross,rotate=45,inner sep=1.5,color=red] at (\xa+\pxa+0.1,\ya+6-\pya-0.1) {};
	\node[draw,thick,cross,inner sep=1.5,color=red] at (\xa+\pya+0.1,\ya+6-\pxa-0.1) {}; 
	
	\node[draw,thick,cross,inner sep=1.5,color=red] at (\xa+\pxa-0.1,\ya+6-\pya+0.1) {};
	\node[draw,thick,cross,rotate=45,inner sep=1.5,color=red] at (\xa+\pya-0.1,\ya+6-\pxa+0.1) {};
	
	\node[draw,thick,circle,inner sep=1.,color=red,fill=white] at (\xa+\pxb,\ya+6-\pyb) {};
	\node[draw,thick,circle,inner sep=1.,fill,color=red] at (\xa+\pyb,\ya+6-\pxb) {};
	\node[draw,thick,circle,inner sep=1.,fill,color=red] at (\xa+\pxc,\ya+6-\pyc) {};
	\node[draw,thick,circle,inner sep=1.,color=red,fill=white] at (\xa+\pyc,\ya+6-\pxc) {};
	\node[draw,thick,circle,inner sep=1.,fill,color=red] at (\xa+\pxd,\ya+6-\pyd) {};
	\node[draw,thick,circle,inner sep=1.,color=red,fill=white] at (\xa+\pyd,\ya+6-\pxd) {};
	
	\draw[->,thin,black] (\xa+\dxa-2,\ya+0.5) -- (\xa+\dxa-1,\ya+0.5);
	\node[black,right,scale=0.75] at (\xa+\dxa,\ya+1) {$x\leftarrow x$};
	\node[black,right,scale=0.75] at (\xa+\dxa,\ya) {$\nu\leftarrow \op{rand\_dir}\of{}$};
			
    \def\ya{-2*\dya};
     \node[left,scale=0.75] at (\xa-3,\ya+6-3) {$x+\hat{\nu}$};
    \foreach \i in {1,...,5} {
        \draw [thin,dotted,gray] (\xa+\i,\ya+1) -- (\xa+\i,\ya+5)  node[solid,black,above] at (\xa+\i,\ya+5.3) {};
    }
    \foreach \i in {1,...,5} {
        \draw [thin,dotted,gray] (\xa+1,\ya+\i) -- (\xa+5,\ya+\i) node[solid,black,left] at (\xa+0.7,\ya+6-\i) {};
    }
    \foreach \i in {1,...,5} {
      \draw [thin,black!100] (\xa+5,\ya+6-\i) -- (\xa+\i,\ya+6-\i) -- (\xa+\i,\ya+6-1);
    }
    \draw [thin,dashed,black] (\xa+1,\ya+5) -- (\xa+5,\ya+1);
    \def\pxa{4};
    \def\pya{3};
    \def\pxb{1};
    \def\pyb{3};
    \def\pxc{2};
    \def\pyc{3};
    \def\pxd{1};
    \def\pyd{2};
    \node[left,scale=0.65] at (\xa+0.75,\ya+6-\pya) {$b=a_{0}$};
    \draw [thin,black] (\xa+0.85,\ya+6-\pya) -- (\xa+1,\ya+6-\pya);
    \node[left,scale=0.65] at (\xa+0.75,\ya+6-\pxa) {$b_{0}$};
    \draw [thin,black] (\xa+0.85,\ya+6-\pxa) -- (\xa+1,\ya+6-\pxa);
    \node[left,scale=0.65] at (\xa+0.75,\ya+6-\pxb) {$a$};
    \draw [thin,black] (\xa+0.85,\ya+6-\pxb) -- (\xa+1,\ya+6-\pxb);

    \draw [thin,dotted,red] (\xa+\pxb,\ya+6-\pyb) -- (\xa+\pyb,\ya+6-\pxb);
    \draw [thin,dotted,red] (\xa+\pxc,\ya+6-\pyc) -- (\xa+\pyc,\ya+6-\pxc);
    \draw [thin,dotted,red] (\xa+\pxd,\ya+6-\pyd) -- (\xa+\pyd,\ya+6-\pxd);
	\node[draw,thick,circle,inner sep=1.,color=red,fill=white] at (\xa+\pxb,\ya+6-\pyb) {};
	\node[draw,thick,circle,inner sep=1.,fill,color=red] at (\xa+\pyb,\ya+6-\pxb) {};
	\node[draw,thick,circle,inner sep=1.,fill,color=red] at (\xa+\pxc,\ya+6-\pyc) {};
	\node[draw,thick,circle,inner sep=1.,color=red,fill=white] at (\xa+\pyc,\ya+6-\pxc) {};
	\node[draw,thick,circle,inner sep=1.,fill,color=red] at (\xa+\pxd,\ya+6-\pyd) {};
	\node[draw,thick,circle,inner sep=1.,color=red,fill=white] at (\xa+\pyd,\ya+6-\pxd) {};

    \def\xa{-1};
    \def\ya{-0.5*\dya+2.5};
    \def\xb{\xa+18};
    \def\yb{\ya-10};
    \node[below right,scale=0.85] at (\xa+0.4,\ya-0.4) {legend:};
    \draw[color=black,opacity=0.1] (\xa,\ya) -- (\xb,\ya) -- (\xb,\yb) -- (\xa,\yb) -- cycle;


    \def\xa{3};
    \def\ya{-0.9*\dya};
    \foreach \i in {1,...,3} {
        \draw [thin,dotted,gray] (\xa+\i,\ya+1) -- (\xa+\i,\ya+3) node[solid,black,above] at (\xa+\i,\ya+3.3) {};
    }
    \foreach \i in {1,...,3} {
        \draw [thin,dotted,gray] (\xa+1,\ya+\i) -- (\xa+3,\ya+\i) node[solid,black,left] at (\xa+0.7,\ya+6-\i) {};
    }
    \foreach \i in {1,...,3} {
      \draw [thin,black!100] (\xa+3,\ya+4-\i) -- (\xa+\i,\ya+4-\i) -- (\xa+\i,\ya+4-1);
    }
    \draw [thin,dashed,black] (\xa+1,\ya+3) -- (\xa+3,\ya+1);

    \node[left,scale=0.75] at (\xa-1,\ya+4-2) {$x$};
    \def\pxa{3};
    \def\pya{2};
    \node[left,scale=0.65] at (\xa+0.75,\ya+4-\pya) {$a$};
    \draw [thin,black] (\xa+0.85,\ya+4-\pya) -- (\xa+1,\ya+4-\pya);
    \node[left,scale=0.65] at (\xa+0.75,\ya+4-\pxa) {$b$};
    \draw [thin,black] (\xa+0.85,\ya+4-\pxa) -- (\xa+1,\ya+4-\pxa);
    \draw [thin,dotted,red] (\xa+\pxa,\ya+4-\pya) -- (\xa+\pya,\ya+4-\pxa);
	\node[draw,thick,cross,rotate=45,inner sep=1.5,color=red] at (\xa+\pxa,\ya+4-\pya) {};
	\node[draw,thick,cross,inner sep=1.5,color=red] at (\xa+\pya,\ya+4-\pxa) {}; 
    \draw[<->,black] (\xa+4.25,\ya+2) -- (\xa+5.75,\ya+2) node[scale=0.75,right] at (\xa+6.5,\ya+2) {$z_{a}\of{x} \obar{z}_{b}\of{x}\ $ (external)};

    \def\ya{-1.5*\dya};
    \foreach \i in {1,...,3} {
        \draw [thin,dotted,gray] (\xa+\i,\ya+1) -- (\xa+\i,\ya+3) node[solid,black,above] at (\xa+\i,\ya+3.3) {};
    }
    \foreach \i in {1,...,3} {
        \draw [thin,dotted,gray] (\xa+1,\ya+\i) -- (\xa+3,\ya+\i) node[solid,black,left] at (\xa+0.7,\ya+6-\i) {};
    }
    \foreach \i in {1,...,3} {
      \draw [thin,black!100] (\xa+3,\ya+4-\i) -- (\xa+\i,\ya+4-\i) -- (\xa+\i,\ya+4-1);
    }
    \draw [thin,dashed,black] (\xa+1,\ya+3) -- (\xa+3,\ya+1);

    \node[left,scale=0.75] at (\xa-1,\ya+4-2) {$x$};    
    \def\pxa{3};
    \def\pya{2};
    \node[left,scale=0.65] at (\xa+0.75,\ya+4-\pya) {$a$};
    \draw [thin,black] (\xa+0.85,\ya+4-\pya) -- (\xa+1,\ya+4-\pya);
    \node[left,scale=0.65] at (\xa+0.75,\ya+4-\pxa) {$b$};
    \draw [thin,black] (\xa+0.85,\ya+4-\pxa) -- (\xa+1,\ya+4-\pxa);
    \draw [thin,dotted,red] (\xa+\pxa,\ya+4-\pya) -- (\xa+\pya,\ya+4-\pxa);
	\node[draw,thick,circle,inner sep=1.,color=red,fill=white] at (\xa+\pxa,\ya+4-\pya) {};	
	\node[draw,thick,circle,inner sep=1.,color=red,fill=red] at (\xa+\pya,\ya+4-\pxa) {};

    \def\ya{-2*\dya};
    \foreach \i in {1,...,3} {
        \draw [thin,dotted,gray] (\xa+\i,\ya+1) -- (\xa+\i,\ya+3) node[solid,black,above] at (\xa+\i,\ya+3.3) {};
    }
    \foreach \i in {1,...,3} {
        \draw [thin,dotted,gray] (\xa+1,\ya+\i) -- (\xa+3,\ya+\i) node[solid,black,left] at (\xa+0.7,\ya+6-\i) {};
    }
    \foreach \i in {1,...,3} {
      \draw [thin,black!100] (\xa+3,\ya+4-\i) -- (\xa+\i,\ya+4-\i) -- (\xa+\i,\ya+4-1);
    }
    \draw [thin,dashed,black] (\xa+1,\ya+3) -- (\xa+3,\ya+1);

    \node[left,scale=0.75] at (\xa-1,\ya+4-2) {$x+\hat{\nu}$};    
    \def\pxa{3};
    \def\pya{2};
    \node[left,scale=0.65] at (\xa+0.75,\ya+4-\pya) {$a$};
    \draw [thin,black] (\xa+0.85,\ya+4-\pya) -- (\xa+1,\ya+4-\pya);
    \node[left,scale=0.65] at (\xa+0.75,\ya+4-\pxa) {$b$};
    \draw [thin,black] (\xa+0.85,\ya+4-\pxa) -- (\xa+1,\ya+4-\pxa);
    \draw [thin,dotted,red] (\xa+\pxa,\ya+4-\pya) -- (\xa+\pya,\ya+4-\pxa);
	\node[draw,thick,circle,inner sep=1.,color=red,fill=red] at (\xa+\pxa,\ya+4-\pya) {};	
	\node[draw,thick,circle,inner sep=1.,color=red,fill=white] at (\xa+\pya,\ya+4-\pxa) {};

    \draw [decorate,decoration={brace,amplitude=5pt,mirror,raise=2pt},yshift=0pt]
(\xa+3.3,\ya+1) -- (\xa+3.3,\ya+5.5);
	
    \draw[<->,black] (\xa+4.25,\ya+3.25) -- (\xa+5.75,\ya+3.25) node[scale=0.75,right] at (\xa+6.5,\ya+3.25) {$k_{x,\nu}^{a b}\to k_{x,\nu}^{a b}+1$};	
  \end{scope}
\end{tikzpicture}}

%% file: worm_algorithm_for_lattice_CPN_model.bbl
\begin{thebibliography}{99}\itemsep 1pt
  \bibitem{RindlisbacherCPN} T. Rindlisbacher, P. de Forcrand, \emph{Worm Algorithm for the $\CPn{N-1}$ model}, http://dx.doi.org/10.1016/j.nuclphysb.2017.02.021, \emph{arXiv:1610.01435 [hep-lat]} .
  \bibitem{Cremmer} E. Cremmer, J. Schrek, \emph{The supersymmetric non-linear $\sigma$-model in four dimensions and its coupling to supergravity}, Phys. Lett. 74B (1978) 341 .
  \bibitem{Golo} V. L. Golo, A. M. Perelomov, \emph{Solution of the duality equations for the two-dimensional $\SU{N}$-invariant chiral model}, Phys. Lett. 79B (1978) 112 .
  \bibitem{Adda} D. D'Adda, M. L\"uscher, P. Di Vecchia, \emph{A 1/N expandable series of non-linear $\sigma$-models with instantons}, Nucl. Phys. B146 (1978) 63-76 .
  \bibitem{Witten1} E. Witten, \emph{Instantons, the quark model, and the 1/N expansion}, Nucl. Phys. B149 (1979) 285-320 .
  \bibitem{Vecchia} P. Di Vecchia, A. Holtkamp, R. Musto, F. Nicodemi, R. Pettorino, \emph{Lattice $\CPn{N-1}$ Models and their large-N behaviour}, Nucl. Phys. B190 (1981) 719 .
  \bibitem{Vecchia2} P. Di Vecchia, R. Musto, F. Nicodemi, R. Pettorino, P. Rossi, \emph{The Transition from the Lattice to the Continuum: $\CPn{N-1}$ Models at Large N}, Nucl. Phys. B235 (1984) 478-520 .
  \bibitem{Campostrini} M. Campostrini, P. Rossi, and E. Vicari, \emph{Monte Carlo simulation of $\CPn{N-1}$ models}, Phys.Rev. D46 (1992) 2647-2662 .
  \bibitem{Jansen} K. Jansen, U.-J. Wiese, \emph{Cluster algorithms and scaling in $\CP{3}$ and $\CP{4}$ models}, Nucl. Phys. B 370 (1992) 762-772 .
  \bibitem{Rabinovici} E. Rabinovici, S. Samuel, \emph{The $\CPn{N-1}$ model: a strong coupling lattice approach}, Phys. Lett. 101B (1981) 323-326 .
  \bibitem{Chandrasekharan} S. Chandrasekharan, \emph{A new computational approach to lattice quantum field theories}, PoS LATTICE 2008 (2008) 003 .
  \bibitem{Vetter} R. Vetter, \emph{The Worm Algorithm for the $\CPn{N-1}$ Model}, Semester Thesis at ETHZ (2011) .
  \bibitem{Wolff} U. Wolff, \emph{Simulating the All-Order Strong Coupling Expansion IV: CP(N-1) as a loop model}, Nucl. Phys. B832 (2010) 520, \emph{arXiv:1001.2231 [hep-lat]} .
  \bibitem{Prokofev} N. Prokof'ev, B. Svistunov, \emph{Worm Algorithm for Classical Statistical Models}, Phys. Rev. Lett. 87 (2001) 160601, \emph{arXiv:cond-mat/0103146} .
  \bibitem{Gattringer1} F. Bruckmann, C. Gattringer, T. Kloiber, T. Sulejmanpasic, \emph{Dual lattice representation for $\On{N}$ and $\CP{N-1}$ models with a chemical potential}, Phys. Lett. B749 (2015) 495-501, \emph{arXiv:1507.04253 [hep-lat]} .
  \bibitem{Flynn} J. Flynn, A. J\"uttner, A. Lawson, F. Sanfilippo, \emph{Precision study of critical slowing down in lattice simulations of the $\CPn{N-1}$ model}, arXiv:1504.06292 [het-lat] .
\end{thebibliography}
